\documentclass[runningheads]{llncs}
\usepackage{cite}
\usepackage{graphicx}
\usepackage{listings}
\usepackage{color}
\usepackage[caption=false]{subfig}
\usepackage{courier}
\usepackage{bbding}
 \usepackage[table,xcdraw]{xcolor}
 \usepackage{multirow}
\begin{document}

\title{Accelerating Deep Learning Inference with Cross-Layer Data Reuse on GPUs}

\titlerunning{Accelerating Deep Learning Inference with Cross-Layer Data Reuse on GPUs}
%
\author{Xueying Wang\inst{1,2}
\and
Guangli Li\inst{1,2}
\and
Xiao Dong\inst{1,2}
\and
Jiansong Li\inst{1,2}
\and \\
Lei Liu\inst{1(}\Envelope\inst{)} 
\and
Xiaobing Feng\inst{1,2}
}
\authorrunning{X. Wang et al.}
%
\institute{State Key Laboratory of Computer Architecture, \\Institute of Computing Technology, Chinese Academy of Sciences, China \and
School of Computer Science and Technology, \\University of Chinese Academy of Sciences, China \\
\email{\{wangxueying,liguangli,dongxiao,lijiansong,liulei,fxb\}@ict.ac.cn}}
\maketitle              
\begin{abstract}
Accelerating the deep learning inference is very important for real-time applications.
In this paper, we propose a novel method to fuse the layers of convolutional neural networks (CNNs) on Graphics Processing Units (GPUs), which applies data reuse analysis and access optimization in different levels of the memory hierarchy.
To achieve the balance between computation and memory access, we explore the fusion opportunities in the CNN computation graph and propose three fusion modes of convolutional neural networks: straight, merge and split.
Then, an approach for generating efficient fused code is designed, which goes deeper in multi-level memory usage for cross-layer data reuse.
The effectiveness of our method is evaluated with the network layers from state-of-the-art CNNs on two different GPU platforms, NVIDIA TITAN Xp and Tesla P4. 
The experiments show that the average speedup is 2.02 $\times$ on representative structures of CNNs, and 1.57$\times$ on end-to-end inference of SqueezeNet.
\end{abstract}
\section{Introduction}
Convolutional neural networks (CNNs) have become more and more popular in deep learning applications, including image classification and video recognition. 
For modern heterogeneous parallel computing platforms such as Graphics Processing Units (GPUs), there has been a rising interest in efficient implementation of deep learning systems.
There are several kinds of operators in deep neural networks, such as convolution, batch normalization, and activation. 
Generally, GPU-based deep learning systems launch kernels for a single operation many times, which may cause extra data transmission overheads.
Complex computation tasks are usually bounded by arithmetic bandwidth and large-scale data transmission are bounded with memory bandwidth.
The bottleneck of executing kernel varies depending on the applications and GPU devices.
For pooling, activation and some kind of convolution operations with small size, 
the workloads are limited by the transmission speed of memory access.

CNN architectures are going deeper and have become too complicated to infer in real-time systems.
The increasing size of deep CNNs demands more on computing systems and  GPUs provide the primary computation for CNN applications. 
However, the performance of CNN inference is subject to computation and memory bandwidth constraints.
There is an increasing gap between memory bandwidth and computing performance on emerging GPUs.

Meanwhile, CNNs are tending to be very deep, such as GoogLeNet~\cite{szegedy2015going}, and usually consist of dozens or hundreds of layers.
Some novel architectures, such as inception and residual connections, resulting in deeper and wider neural networks.
For accelerating the inference, some light-weight and efficient CNNs are proposed, such as SqueezeNet\cite{SqueezeNet} and MobileNet\cite{howard2017mobilenets}.

The inference systems are usually parallel and have hierarchical memory and the memory access bandwidth is the potential bottleneck for accelerating neural networks. 
In the architectural design of the GPU, the latency of global memory is much higher than shared memory.
New GPU architectures are emerging, Volta, Turing and Ampere. However, the new hardware architectures are focused on the single layer execution time rather than the data reuse across layers in different memory level.

As such, reusing shared memory data can achieve much more performance improvement besides the benefits of hardware upgrades.
Inspired by kernel fusion~\cite{KernelFusionCGO19',filipovivc2015optimizing,wu2012optimizing,wahib2014scalable},  we propose a cross-layer data reuse approach by fusing kernels to increase the data locality and reuse efficiency cross the layers. 

\begin{figure}
	\centering
	\includegraphics[width=0.6\linewidth]{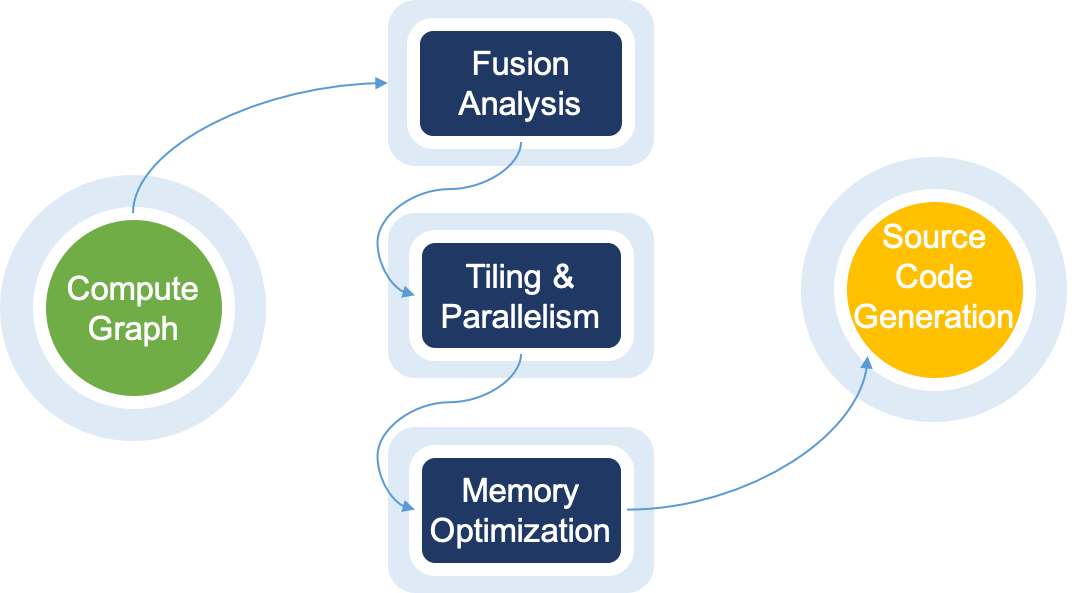}
	\caption{Workflow of Our Cross-layer Data Reuse Method}
	\label{fig:screenshot001}
\end{figure}
Unfortunately, few works have addressed the issue of how to formally describe and fuse deep CNNs across layers in detail.
In particular, the performance of our method can catch up with the existing acceleration library.
Our goal is to develop a strategy for generating high performance code of deep CNN applications by exploring the cross-layer data reuse.
We design a cross-layer data reuse optimization method, which inputs the compute graph of CNN layers and generates the source code for GPUs (Fig.~\ref{fig:screenshot001}).
The fusion strategies include analyzing the input graph for fusion, tiling the data and parallelism on devices and optimizing the memory usage on multi-level memory hierarchy. 

The main contributions of this paper are:
\begin{itemize}
\item[-] To find more optimization opportunities for subsequent fusing, we characterize the computational procedure in CNNs and summarized three fusion modes (straight, merge and split) formally.
\item[-] We propose a fusion method that can reuse on-chip memory by making full use of multi-layer memory on GPUs.
Based on the method, we build a code generator, which can automatically generate a high-performance fused kernel according to determined fusion mode.

\item[-] We conduct experiments on representative networks and analyze the results.
The experimental results show that the performance of our method outperforms the GPU-accelerated deep neural network library, cuDNN~\cite{chetlur2014cudnn}.
\end{itemize}
\section{Hierarchy of Modern GPUs}
In this section, we first introduce the memory hierarchy for modern GPU architectures, which is the basis for CNN application optimization.
Then, we give a motivating example and describe the data reuse methodology in convolution applications.

\subsection{Hierarchy of GPUs and CUDA} 
Compute Unified Device Architecture (CUDA) is a parallel computing platform and programming model for GPUs~\cite{nvidia2007compute}, which exposes programmers to the concepts of memory hierarchy and threads hierarchy~\cite{ProCUDA}.
Accelerating deep learning performance on complex memory hierarchy needs to make full use of memory units and compute units.

\begin{figure}
	\centering
	\includegraphics[width=0.4\linewidth]{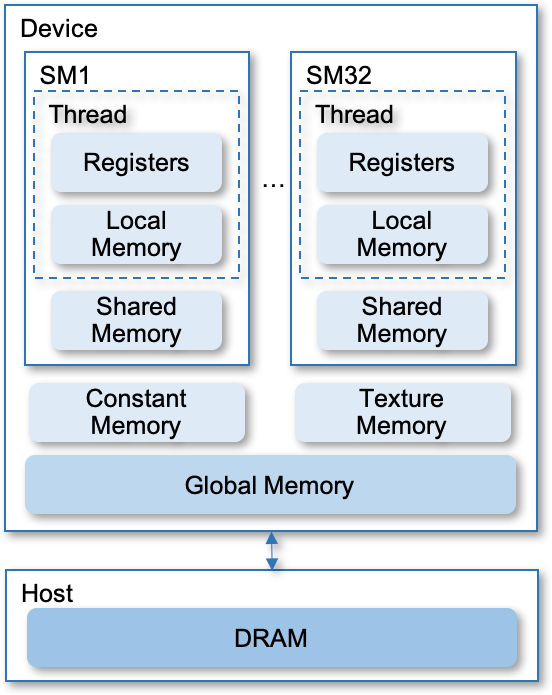}
	\caption{GPU Memory Hierarchy}
	\label{fig:screenshot002}
\end{figure}

As shown in Fig.~\ref{fig:screenshot002}, there are many programmable memories at different levels of GPU devices.
GPU memory units vary from access pattern to management. 
Modern GPUs contain a lot of Stream Multiprocessors (SMs) which are parallel executed on the board.
Each SM has its shared memory, which can be accessed by threads in the same block. 
Multiple blocks can be launched on the SM, but each block can only access its private shared memory.
Registers and local memory can only be visited by a single thread. 
If the size of the required resisters is larger than the size each thread allocated, local memory will be used.
Constant memory, texture memory, and global memory can be visited by all threads.
Constant memory is a kind of read-only memory, which needs to be transferred to GPU device memory from CPU memory before launching the kernel.
Texture memory is read-only and optimized for 2D access.
Generally, data will be prepared by copying memory data from host memory to global memory before the kernel launched.

On-chip memory is fast and close to chips while off-chip memory is slow and far away from chips.
Different types of memory have different access patterns.
Registers and local memory are both private to each thread. 
But registers are on-chip and have low latency and local memory is off-chip and has high latency.  
Shared memory is organized by equal-sized banks. 
Accessing data in the same bank simultaneously will cause shared memory bank conflict and get higher latency.
The global memory is off-chip and large memory capacity, but also has high access latency.
The average latency is about 7000$\times$ higher than register latency and 5$\times$ higher than shared memory latency~\cite{fang2018benchmarking}.

GPUs have become the most popular accelerator with high computational throughput.  
Large and deep neural networks require substantial computing and memory throughput and existing methods do not make good use of this multi-level memory hierarchy for the complex architecture of GPUs.

\subsection{Motivating Example}
Convolution operation is the most time-consuming part of the whole neural network.
Convolution, pooling, activation, element-wise concatenation and addition are basic operations and layers in recent neural networks.
Although the deep convolutional operations are compute-bound, the pooling, activation, element-wise operation and convolution layers with small input channels are memory-bound.
This requires a mechanism to achieve a balance between computation and memory access.

\begin{figure}[htp]
	\centering
	\subfloat[Original\label{fig:submotivatinga}]{%
		\includegraphics[height=0.25\textwidth]{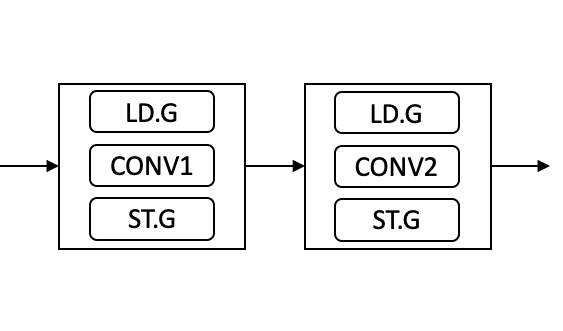}%
	}\hfil
	\subfloat[Fused\label{fig:submotivatingb}]{%
	\includegraphics[height=0.25\textwidth]{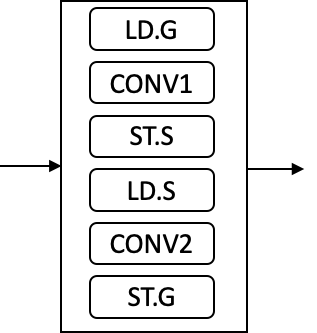}%
	}	
	\caption{CNN Motivating Example}
	\label{fig:motivating}
	
\end{figure}
The benefit of cross-layer data reuse on two CNN layers is the difference in latency and throughput between shared memory and global memory.
The original and fused main kernel structures are shown in Fig.~\ref{fig:motivating}.
\texttt{LD.G} and \texttt{ST.G} illustrate global memory data load and store instructions.
\texttt{LD.S} and  \texttt{ST.S} are the load and store instructions, which read and write on shared memory.
\texttt{CONV1} and  \texttt{CONV2} are the computation of the first and the second convolution layers.
Figure~\ref{fig:submotivatinga} depicts the original kernels, which individually compute two convolutional operators. Each kernel loads the input data from global memory and stores the result to global memory, which implies twice execution of \texttt{LD.G} and \texttt{ST.G}.
As shown in the FIg.~\ref{fig:submotivatingb}, one fused convolutional kernel only contain once \texttt{LD.G} and \texttt{ST.G}, and use the \texttt{ST.S} and \texttt{LD.S} to buffer the intermediate data.

Each layer may fetch data from off-chip memory, compute in on-chip memory and store to off-chip memory.
But fused convolution layers can reduce the off-chip global memory read/store transactions between two layers. We load data from shared memory and store data in shared memory, which means converts the global memory load/store to the shared memory load/store.

\section{Method}
In this section, the method of fusing convolutional layers on GPUs is depicted in detail.
First, we analyze the fusion optimization opportunities of diverse convolution neural networks and sum up three typical fusion modes.
In the second step, we use the data dependency to determines the size of the redundant data on each SM and the size of the tile, which takes the relationship between the input CNN layer and the sequence layer into consideration.
Finally, the use of multi-level memory on the device is optimized during the parallel code generation phase.

\subsection{Fusion Mode Formulation }
\begin{figure}[htp]
	\centering
	\subfloat[Straight\label{fig:subim1}]{%
		\includegraphics[height=0.25\textwidth]{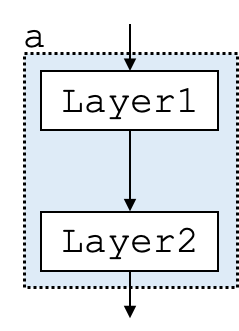}%
	}\hfil
	\subfloat[Split\label{fig:subim2}]{%
	\includegraphics[height=0.25\textwidth]{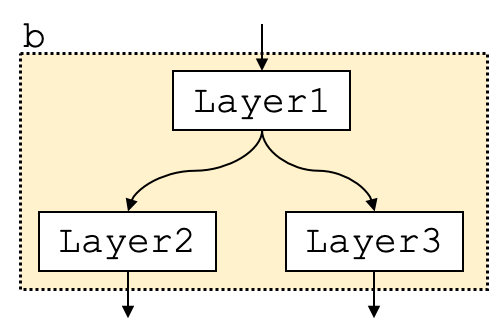}%
	}\hfil
	\subfloat[Merge\label{fig:subim3}]{%
		\includegraphics[height=0.25\textwidth]{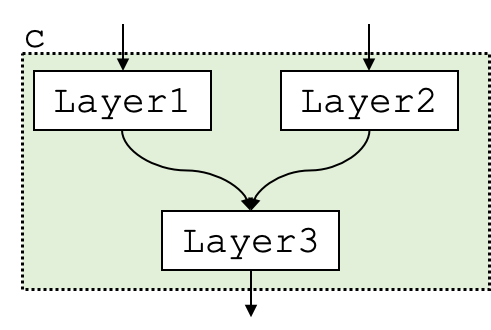}%
	}	
	\caption{Different Fusion Mode}
	\label{fig:screenshot010}
\end{figure}

The neural network architecture is constantly changing and it is necessary to formalize some common architectures for neural network fusion, similar to the hierarchical representations~\cite{liu2017hierarchical} in Neural Architectural Search.

Subject to the capacity of shared memory and high latency caused by bank conflict, the cross-layer data can not be stored in on-chip memory and reused on more than two layers.
Using too much shared memory resources will cause the high access latency for shared memory bank conflict, which may cause performance decrease.

To conclude the common layer architectures in the convolutional neural networks, we propose three basic fusion modes.
As described in Fig.~\ref{fig:screenshot010}, the cross-layer relationships are summarized to three fusion modes. Fig.~\ref{fig:subim1} describes the straight fusion mode, which makes the output data of \texttt{Layer1} reuse for \texttt{Layer2}.
Fig.~\ref{fig:subim2} gives a split mode, which \texttt{Layer1} can be the input of both \texttt{Layer2} and \texttt{Layer3}.
Fig.~\ref{fig:subim3} is a merge mode that has two layers as the input of the third layer, which suggests that \texttt{Layer3} needs the correct computation results of \texttt{Layer1} and \texttt{Layer2}.

These three basic modes can be widely found in most deep neural networks. 
For example, neural networks with sequential layers (rather than residual and inception structures) are ubiquitous, which can be divide into mode (a) Straight.
Residual module and inception connection make the network wider, deeper and more complicated. There are a variety of mixed fusion modes in such neural network architectures, which brings challenge to cross-layer data reuse analysis.

\begin{figure}[htp]
	\centering
	\subfloat[Inception\label{fig:inception}]{%
		\includegraphics[height=0.4\textwidth]{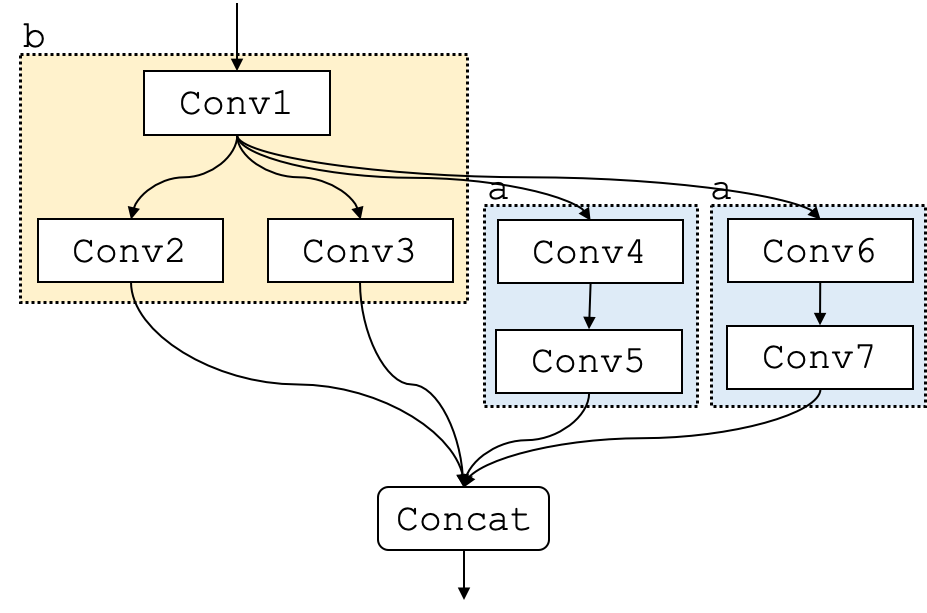}%
	}\hfil
	\subfloat[Residual\label{fig:residual}]{%
		\includegraphics[height=0.4\textwidth]{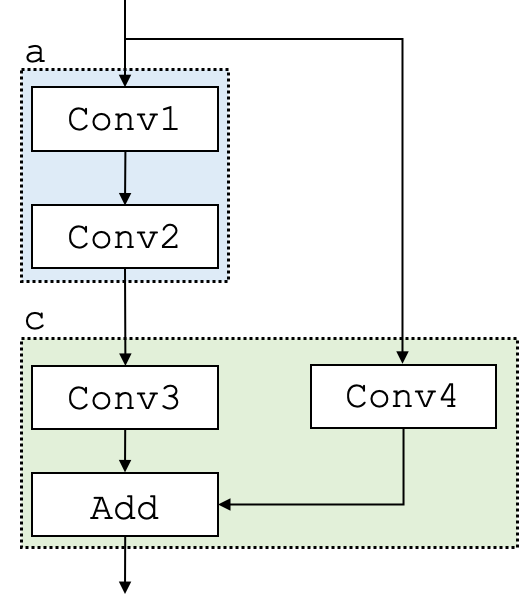}%
	}	
	\caption{Fusion Example for Inception and Residual Neural Networks}
	\label{fig:fusionexample}
\end{figure}

Figure~\ref{fig:fusionexample} abstracts the convolution layers in the network (some other operations, such as ReLU and pool, are omitted).
In this figure, the boxes represent layers, arrows represent the dataflow dependency, dotted boxes represent fusion blocks and the letter upon the dotted box represents fusion mode.

In Fig.~\ref{fig:inception}, the inception module for fusion strategy is depicted, which includes two modes, mode \texttt{a} and mode \texttt{b}.
In the mode \texttt{b} block, the output data of \texttt{Conv1} can reuse and input the \texttt{Conv2} and \texttt{Conv3}.     
As shown in Fig.~\ref{fig:residual}, the residual connection is divided into three fusion blocks. The block who belongs to mode \texttt{a} contains \texttt{Conv1} and \texttt{Conv2}, that the result of \texttt{Conv1} will be reuse. The block which is mode \texttt{c} means the \texttt{Add} operations can reuse the results of \texttt{Conv3} and \texttt{Conv4} on-chip.  

\subsection{Tiling and Parallelism} 
Unifying CNN layers into a single kernel is a challenge for layer fusion because of the different data size and filter shape diversity.
Tiling is an important parallel strategy on GPU programming.
The fundamental problem for layer fusion is how to tile the data on the parallel system with multi-level memory hierarchy, which called hierarchy overlapped tiling~\cite{zhou2012hierarchical}.
\begin{figure}
	\centering
	\includegraphics[width=0.9\linewidth]{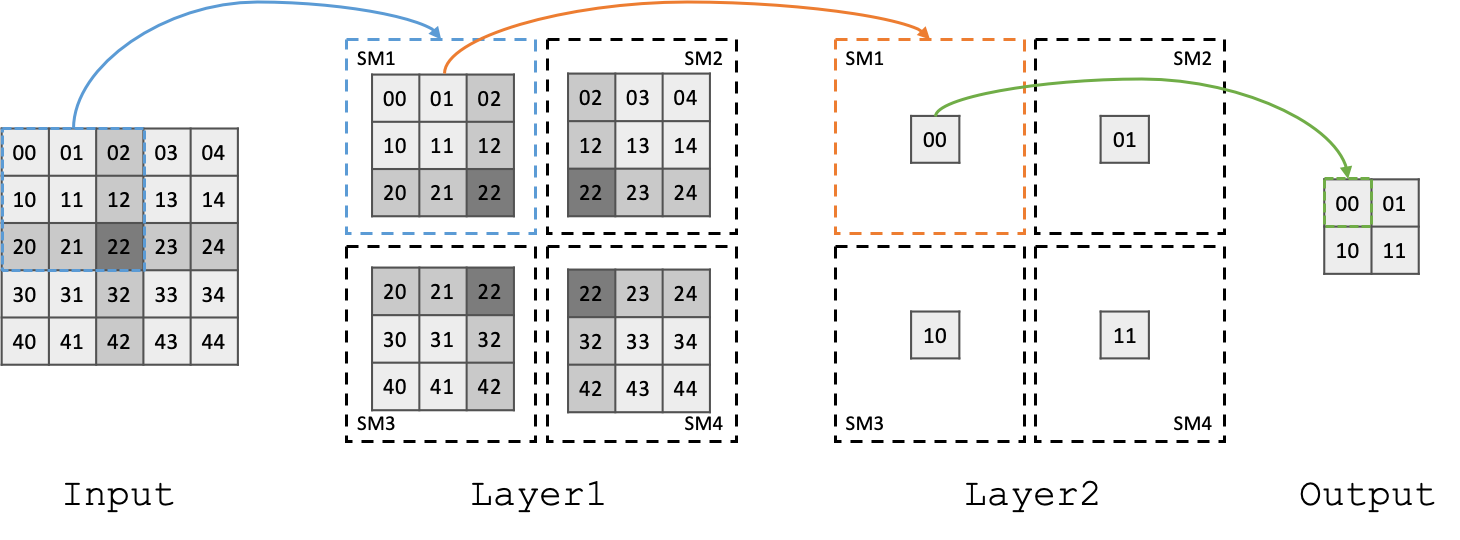}
	\caption{Tiling and Parallelism Example}
	\label{fig:tilingexample}
\end{figure}

Our tiling strategy is to tile each output images and feature maps into small tiles on the dimension height and width, and implement implicit General Matrix Multiplication (GEMM) convolution algorithm.
Each single output pixel depends on all input pixels through all channels within the window of filter, and the convolution operations whose filter height and width are larger than $1\times1$ need redundant computation and data storage.
There is no need to pay additional attention to element-wise operations because of data independency.

Considering data reuse across layers, the parallel model for the fused layers is restricted by the CNN layer parameters.
The filter with large size, which is larger than 1, will cause redundant computing for data dependence.
Figure~\ref{fig:tilingexample} shows a tiling example for two fused neural network layers computation on 4 SMs. 
The SMs are parallel on GPUs, and shared memory of each block is isolate and private, which means that redundant data storage is necessary for on-chip data reuse.
The center data replicates in each SM and the border data around the center in \texttt{Layer1} are redundant in adjacent SM, which makes it be available for the subsequent \texttt{Layer2}.
The tiling size will greatly affect the performance, too large or too small will cause performance degradation.
The small tiling size will result in too much data redundant computation and reduce the earning of data reuse. 
As shown in Fig.~\ref{fig:tilingexample}, tiling size of $3\times3$ will make 36 elements stored on the GPUs while the input size is 25 if the convolution filter size is 3.
The tiling size of one will not cause any redundant data, which also means that the convolutional filter size will influence the tiling strategy.

However, the larger tiling size is not always better. There is a tradeoff between choosing large size and small size.
The large tiling size will occupy more computing resources, which reduces the degree of parallelism of the computation.
As the intermediate data storage location, shared memory is an important bottleneck resource for the cross-layer data reuse.
More shared memory are allocated by each thread block, fewer thread blocks can be processed simultaneously by an SM.
If shared memory or register is unavailable to process at least one block on each SM, the kernel will fail for resource limitation.
In the worst case, the number of parallel blocks on an SM is reduced to 1 and the latency cannot be hidden.
Therefore, it is recommended to use less than 1/3 shared memory to achieve high performance~\cite{ProCUDA}.

To this end, we design a simple tunning tool to find a relatively optimal tiling size.
The tunning tool only searches for the combination of the common factors of the output layer.
For example, the combination (4,3) means $4\times4$ tiling size and (3,3) grid size. 
For the output size (12,12), the tunning search space will be \{(4,3), (2,6), (3,4), (6,2)\}.
If output height and width are prime, the size larger than the number will be chosen as the basis for tuning.
We allocate each thread one point computation, and if the tiling size is larger than the block size limitation, there will be for loops inside thread across the width and height dimension.

\subsection{Memory Optimization}
Memory bandwidth, compute resources, instruction and memory latency are three common limiters to performance for a kernel.
For CPU programming, it can be safe to ignore the cache line size or the number of registers. 
But for GPUs, a runtime error occurs when the size of the programmable memory requested exceeds the computing resources. Improper memory usage will cause drastic performance degradation.
For this, the memory strategy needs to be carefully considered while programming on GPUs.
 
\paragraph{Shared Memory Usage}
For the optimization of shared memory, reduce bank conflict with memory padding and use synchronization statements to guarantee the correctness of data in memory.

When a warp accessing different words with the address in the same bank, a 32-way bank conflict will occur.
To avoid the shared memory bank conflict, our method generates code with memory padding.
Extra unused shared memory is paid to allocate a redundant column and row for padding, which is effective in reducing bank conflict.
Memory padding strategy will work after only padding on either row or column in most situation, which will not cause too much resource waste than padding both row and column.

To guarantee the correctness of the data in shared memory, explicit barrier and memory fences are necessary.
\texttt{\_\_syncthreads()} is intra-block synchronization in CUDA, which is used in each thread block to ensure that all threads writing partial results to shared memory have completed before any threads read the final results.
Threads in a block will wait for all threads to execute this instruction, which maintains correctness of data in memory with less performance loss.

\paragraph{Read-only Data Optimization}
Constant memory and read-only cache are optimized for read-only data accessing, which can speed up the data load efficiently.
Constant memory has restricted size, which usually is 64 KB, and read-only cache is 48KB.
The unified L1/texture cache is read-only cache, which is an alternative to L2 cache when accessing read-only data in global memory.

The input data of the first layer and the filter weight of the convolution layers are read-only and do not need to write back during computation.
The strategy for our layer fusion cases are using global memory with read-only cache for input data and using constant memory for filter and bias data. 
If the size of filter and bias data exceeds the constant memory restriction, global memory with read only cache will be used.

\paragraph{Padding Strategy}
Obviously, these convolution layers are not aligned, especially when padding and stride operations exist.
There are two alternatives for padding operations, using branch statement or fill in extra data to the margin. 
Because there are no branch prediction mechanisms on GPUs, flow control constructs, such as \texttt{if} and \texttt{else} clause, will cause great penalty performance.
Extra data movement will lead to memory bandwidth bottleneck.  

Padding operation widely exists in CNNs, for keeping the output size consistent with the input size.
For layer fusion, we need to conduct padding operation on shared memory within a kernel execution. 
After the first layer computation, we preprocess the data with padding so that there is no padding operation in next layer.

\section{Experiments}
In this section, experimental setup and performance results are given. 
The caffe prototxt files are used as the input in the experiments, and the neural network structures are extracted from the deep neural network compute graph.
\subsection{Experimental Setup}
When developing applications on the GPU, the correctness needs to be paid attention to first, and then the performance of the code is improved.
To verify the correctness and compare the performance, we use cuDNN~\cite{chetlur2014cudnn}, one of the most popular deep learning accelerator libraries on GPUs, as a baseline.
The cuDNN is a deep learning library, which is closed-source and NVIDIA hardware limited.
Most of the deep learning frameworks use cuDNN as computational back-ends.
To eliminate the effect of the algorithm, we use the latest and best-performing version, cuDNNv7, as the baseline and a tool to check the correctness of results.
The routine \texttt{cudnnConvolutionBiasActivationForward()} applies a bias and then an activation to the convolutions, which combines these operations into one kernel.
For a fair comparison, we choose the same convolutional algorithm to compare the performance between cuDNN library and our method.
We set \texttt{IMPLICIT\_GEMM} as a convolution algorithm instead of other specially optimized algorithms, which implicitly performs GEMM without actually form the matrix that holds the data.

Three basic modes for fusing convolution layers are shown in Fig.~\ref{fig:screenshot010}, which is concluded from the mainstream neural networks.
Compute graphs extracted from different neural networks are used to perform our fusion optimization method. 
We extract 4 different convolution neural network layers from state-of-the-art networks, GoogleNet~\cite{szegedy2015going}, MobileNet\cite{howard2017mobilenets}, SqueezeNet\cite{SqueezeNet} and ResNet\cite{he2016deep}.
As shown in Table~\ref{tab1},  the ID represents the fusion mode and the test case number. 
Input and output are clarified with the size information, each with shape [Channel, Height, Width]. 
The batch size of input data is set to 1. 
Filter size is depicted by [Output Channel, Input Channel, Filter Height, Filter Width]/padding, stride, group.

\begin{table}
	\centering
	\caption{Convolutional Neural Network Layers in the Fusion Experiment}\label{tab1}
	\begin{tabular}{|c|c|c|c|c|c|}
		\hline
		 \textbf{ID} &  \textbf{Input} 	& \textbf{Filter1 Size} 	& \textbf{Filter2 Size} 	& \textbf{Filter3 Size} 	& \textbf{Output} 		\\
		\hline
		\rowcolor[HTML]{EFEFEF} 
		\textbf{a.1} & [192,28,28] & [16,192,1,1]/0,1,1 	& [32,16,5,5]/2,1,1		& - 	& [32,28,28] 	\\
		\hline
		\textbf{a.2 }& [16,80,80] & [16,1,3,3]/1,1,16 	& [16,1,1,1]/0,1,1	&-		& [16,80,80] 	\\
		\rowcolor[HTML]{EFEFEF} 
		\hline
		\textbf{b.1} & [64,56,56] & [16,64,1,1]/0,1,1	 	& [64,16,1,1]/0,1,1 	& [64,16,3,3]/1,1,1 	& [128,56,56] 	 \\
		\hline
		\textbf{c.1} & [64,56,56] & [256,64,1,1]/0,1,1	& [256,64,1,1]/0,1,1	& [64,256,1,1]/0,1,1		& [64,56,56] 	\\
		\hline
	\end{tabular}
\end{table}

We evaluate the experimental results on two different GPU devices, NVIDIA TITAN Xp and Tesla P4.
TITAN Xp GPU achieves a peak throughput of 12.15 TeraFLOPS, 6074 GB/s shared memory bandwidth and 547.7 GB/s global memory bandwidth with 30 SMs.
P4 GPU is Pascal architecture and has a 5.5 TeraFLOPS single-precision peak performance, 2721 GB/s shared memory bandwidth, 192 GB/s global memory bandwidth with 20 SMs.

The code was compiled using the NVIDIA CUDA compiler (version 10) with flags ‘-O3’.
We execute each kernel over 5 times for run-to-run variation counting and report the average time.
GPU timers are used to collect the time information of the applications.

\subsection{Performance Results and Analysis}
The performance of different applications is often strongly influenced by parallel strategies and memory access performance. 
We implement our layer fusion method with four different neural network architectures and demonstrate the performance on two different GPU devices.
To clarify the effectiveness of the whole neural network with our method, we conduct the fusion method on the SqueezeNet\cite{SqueezeNet}.
The profiling analysis is conducted on the kernels to find out the relationship between our method and cuDNN on memory and computation. 

\begin{figure}
	\centering
	\includegraphics[width=0.8\linewidth]{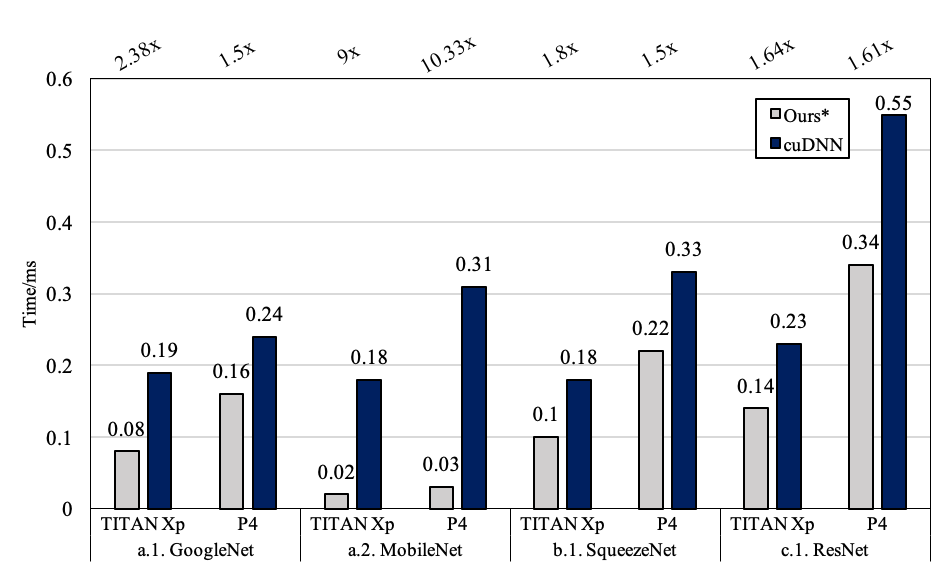}
	\caption{The Experiment Result of Convolutional Neural Network Fusion}
	\label{fig:screenshot009}
\end{figure}
In Fig.~\ref{fig:screenshot009}, we show the speedup of four test cases on TITAN Xp and P4 GPUs.
The left bar of each group is the execution time of fused layers and the right bar is the sum of the execution time of each cuDNN kernels. 
The fusion test cases achieve 1.8$\times$, 9.8$\times$, 1.6$\times$ and 1.62$\times$ speedup.
The average speedup on TITAN Xp is 2.29$\times$ and P4 is 1.91$\times$. 
The experiment a.2 comes from MobileNets and contains depth-wise convolution operations, which called ‘group convolution’ in cuDNN library. 
It calls corresponding grouped convolutional kernels multiple times, which causes performance degradation and 10.33$\times$ speedup on P4 GPU.

\begin{figure}
	\centering
	\includegraphics[width=1\linewidth]{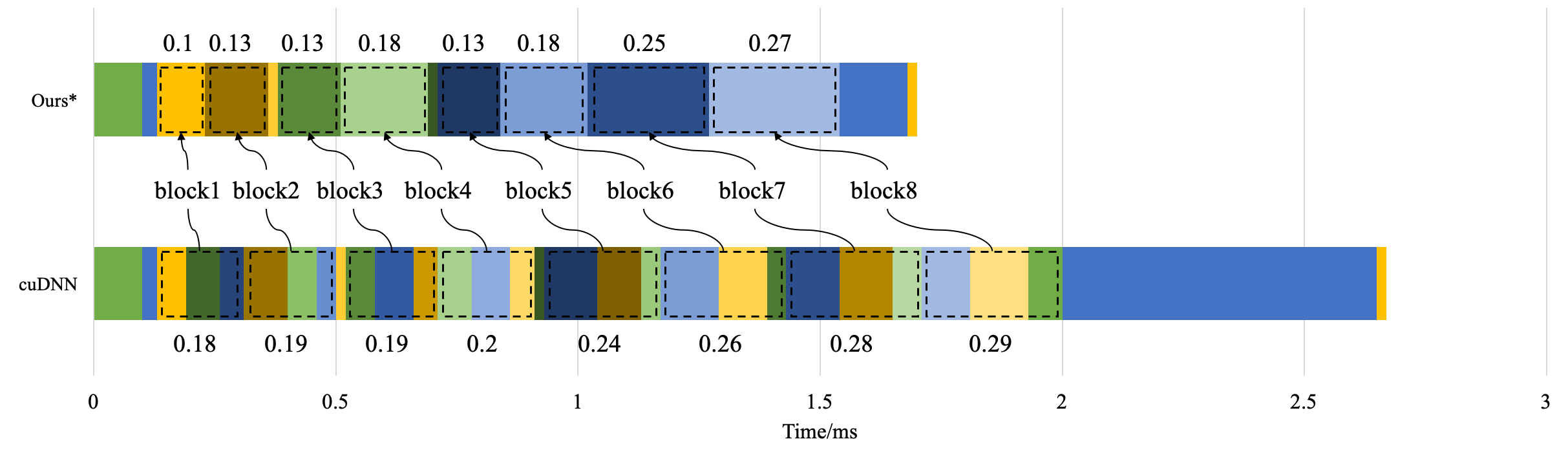}
	\caption{The Experiment Result of SqueezeNet}
	\label{fig:squeezenet}
\end{figure}
We also evaluate our fusion method on the light-weight convolutional neural network, SqueezeNet~\cite{SqueezeNet}.
There are 8 mode \texttt{b} blocks that we can apply our fusion method in this neural networks.
In Fig.~\ref{fig:squeezenet}, we show the execution time of our fused kernels and cuDNN kernels.
The speedup of the whole SqueezeNet on TITAN Xp is 1.57$\times$, and the speedup of the fused blocks to original layers is 1.34$\times$.
The last convolutional layer in the neural network consumes too much more time than the smaller size layers, which is an unusual situation.
For this, we conduct the tiling and parallel strategy of our method on this layers and achieves 4.64$\times$ speedup on this single layer.

\begin{table}[]
	\centering
	\caption{Profiling Metrics on Memory}\label{tab2}
	\begin{tabular}{|c|c|c|c|c|c|c|}
\hline
\multirow{2}{*}{}  & \multicolumn{3}{c|}{ \textbf{Executed Load/Store Instructions}}& \multicolumn{3}{c|}{ \textbf{Global Memory Store Transactions}}     \\ \cline{2-7} 
  & \textbf{Ours*}     & \textbf{cuDNN}     &  \textbf{Ratio}    & \textbf{Ours*}     & \textbf{cuDNN}     &  \textbf{Ratio} \\ \hline
  \rowcolor[HTML]{EFEFEF} 
\textbf{a.1}              & 927472    & 129690 & 7.14:1   & 6272            & 18816     & 1:3             \\ \hline

\textbf{a.2}            & 129600    & 514998 & 1:3.97      & 25600           & 102400    &1:4              \\ \hline
\rowcolor[HTML]{EFEFEF} 
\textbf{b.1}             & 1903552    & 433542 &4.39:1       & 100352          & 225792    & 1:2.25        \\ \hline

\textbf{c.1}            & 2654720    & 1588384 & 1.67:1    & 34008           & 91296     &1:2.68         \\ \hline
\end{tabular}
\end{table}

In Table~\ref{tab2}, the GPU profiling metrics about memory operations are compared between the fused kernels and cuDNN kernels.
The \texttt{ldst\_executed} metric counts the total executed load/store instructions and \texttt{gst\_transactions} gains additional insight into the number of the global memory store transactions. 
The global memory store transactions metric reports the number of coalesced global memory store transactions, which implies the quantity of global memory access and evaluates the global write operations saved by our methods.
The reason why we use the global memory store metric rather than the global memory store metric is that texture memory is utilized as data storage, which will be not counted in the global memory load metrics.
All data need be stored in global memory finally and global memory store transactions metric is much more objective to describe the quantity of global memory transactions.

Our method will introduce redundant computation and also increase the number of load/store instructions but still get an acceleration ratio.
The test case a.2 is group convolution from MobileNet, which is abnormal for executing the same kernel 17 times.
In addition to this structure, we catch 4.4x more load/store instructions execution on devices.
The global memory store transactions ratio between our layer fusion method and the baseline is 1:2.98 on average, while we have much more load/store instructions.
Our method will introduce redundant computation and also load/store instruction but still get a satisfactory acceleration ratio.

\section{Related Work}
To the best of our knowledge, our work is the first about how to generate high performance code by fusing two or more convolutional layers on GPUs, which can achieve competitive performance with the cuDNN library.

Much effort has been made to optimize CNN applications.
Li et al.~\cite{li2016optimizing}  transpose the data and apply different data layouts on different operations to explore the impact of data layouts on the performance of convolutional layers and memory-bound pooling layers.
Data reuse has been explored on fusing CNN layers. Alwani et al. ~\cite{Alwani_2016} proposed pyramid-shaped multi-layer sliding window to handle the input feature maps and verified on FPGA.

Besides the direct code optimization strategy and algorithm, inference framework and DSL are the two main code optimization ways for different parallel devices.
CNN inference framework~\cite{mazaheri2019enhancing} generates Vulkan code and achieve reasonable performance on different platforms.
Halide~\cite{ragan2013halide} is a domain specific language for image processing applications, which introduces the principle of increasing the producer-consumer locality and adopts the loop fusion optimization strategy.
TVM~\cite{chen2018tvm} is a compiler to generate portable code for deep learning applications across diverse hardware platforms.
The source code generated by our method is easy to understand and modify, which is also portable among different GPU platforms.

Kernel fusion is also a hot research point in GPU kernel optimization. 
Wu et.al~\cite{wu2012optimizing} introduce the benefits of kernel fusion in data warehousing applications. 
Wahib et.al\cite{wahib2014scalable} optimize the code with kernel fusion and utilize a heuristic search algorithm for choosing a near-optimal fusion configuration.
The source-to-source compiler~\cite{filipovivc2015optimizing} explores the automatic kernel fusion algorithm for basic linear algebra subprograms routines on GPUs.
Recently, the work of Qiao, B et.al~\cite{KernelFusionCGO19'} depicts kernel fusion problem as finding some cut set of kernels to fuse  in DAG-graph.
Vertices in the graph represent kernels and edges represent the relationship between kernels. 
They provide an algorithm about how to choose kernels while our method provides a method about how to fuse kernels better.

\section{Conclusion}
Considering the characteristics of deep convolution neural networks and underlying GPU architectures, we proposed a cross-layer data reuse method.
The experiments of real-world CNNs show that our method achieves competitive performance and supports the possibility to generate inference code for deep learning applications.

The effectiveness of layer fusion method is evaluated on different test cases and the end-to-end neural network and we get 2.02x and 1.57x speedup on GPUs even with more instructions executed.
We hope that the result of our method will support future research and application about the layer fusion and our method will be widely used and accelerate inference stage for deep learning applications on GPUs. 

\section*{Acknowledgement}
Special thanks to Dr. Zhen Zhang for his helpful discussion about kernel fusion during Xueying Wang's internship.
This work is supported by the National Key R\&D Program of China under Grant No.2017YFB1003103, and the Science Fund for Creative Research Groups of the National Natural Science Foundation of China under Grant No.61521092.

\bibliographystyle{splncs04}
\bibliography{document}

\end{document}